# Parametric analysis of the Lateral Distribution Function of Cherenkov light for Yakutsk EAS Array in the Energy Range 1–20 PeV

**Falah Al-Zubaidi[1], A. A. Al-Rubaiee[1*], B. Hariharan[2,3]**

[1] *Dept. of Physics, College of Science, Mustansiriyah University, Baghdad, Iraq*
[2] *GRAPES-3 Experiment, Cosmic Ray Laboratory, Raj Bhavan, Ooty 643001, India*
[3] *Tata Institute of Fundamental Research, Dr Homi Bhabha Road, Mumbai 400005, India*

**Abstract**

In this research, the simulation of lateral distribution function (LDF) of Cherenkov radiation was performed using CORSIKA software for two hadronic models QGSJET and GHEISHA. This simulation was performed for several elementary particles such as protons, iron nuclei, electrons and gamma quanta, in the range of energies 1-20 PeV for three zenith angles $0°$, $20°$ and $30°$. A parameterization of Cherenkov light LDF was performed for that simulated curves using Lorentzian function. The comparison between the obtained results for LDF of Cherenkov light with that measured with Yakutsk EAS array gave a good agreement within the distances of 100-1000 m from the shower axis.

**Introduction**

Cosmic rays (CRs) are high, ultra-high, and extremely high energy particles of extraterrestrial origin, which constantly impinge the Earth's atmosphere [1]. After the first interaction of a CR particle of very high energy in the atmosphere a multitude of subsequent interactions, leading to particle multiplication, and decay processes give rise to a cascade of secondary particles called extensive air showers (EAS) [2]. EAS develop in a complex way as a combination of electromagnetic cascades and hadronic multiparticle production. It is necessary to perform detailed numerical simulations of air showers to infer the properties of the primary CRs that initiate them. But simulations are a challenge since the number of charged particles in an ultra-high energy shower can be enormous, perhaps exceeding $10^{10}$ [3]. An EAS usually consists of billions of secondary particles, mostly, electrons and muons that arrive at ground level over large areas. The relativistic charged particles in EAS generate Cherenkov light. Above $10^{14}$ eV the only possibility for CR particle detection and measurement is ground based, that is, the detection of one or several of the components of secondary CR. One of the most convenient techniques is the atmospheric Cherenkov technique, that is, the detection of the Cherenkov light in EAS [4].

---

[*]) For correspondence: Email: dr.rubaiee@uomustansiriyah.edu.iq.

In the present work, the simulation of Cherenkov light LDF was performed using CORSIKA code [5, 6] for configuration of Yakutsk EAS array in the energy range (1-20) PeV for different primary CR particles such as (proton, iron nuclei, electron and gamma quanta) and three different zenith angles (0°, 20° and 30°). The LDF curves of Cherenkov light are parameterized using Lorentzian function that gave a new parameters for Cherenkov light LDF.

**The Cherenkov Light LDF**

The total number of Cherenkov photons that radiated by electrons in EAS is directly proportional to the primary energy $E$ [7]:

$$N_\gamma(E) \approx 3.7 \cdot 10^3 \frac{E}{\beta_t} \approx 4.5 \cdot 10^{10} \frac{E}{10^{15} eV} \qquad (1)$$

where $\beta_t$ is the critical energy that is determined as an energy equal to ionization loss of a particle at the $t$-unit: $\beta_t = \beta_{ion} t_0$. For electron, $\beta_{ion} = 2.2 \, MeV \cdot (g \cdot cm^{-2})^{-1}$, $t_0 = 37 \, g \cdot cm^{-2}$ and $\beta_t = 81.4$ MeV [8]. The experimental measurement of this magnitude is rather difficult, therefore one can use the density of Cherenkov radiation LDF- the number of photons per unit of a detector area, which appears as a function of an energy and distance from the shower axis:

$$Q_{(E,R)} = \frac{\Delta N_{\gamma(E,R)}}{\Delta S} \qquad (2)$$

Direct measurements of Cherenkov light showed that the fluctuation of LDF in EAS is essentially less than that of the total number of photons $N_\gamma$ [7, 9]. For parameterization of simulated Cherenkov light LDF, it was used the Lorentzian function, which gave a new four parameters for Cherenkov light LDF as a function of the distance from the shower axis, which is given as:

$$Q(R) = \eta + \frac{2\alpha}{\pi} \frac{\delta}{4(R-\zeta_c)^2 + \delta^2} \qquad (3)$$

Where $\eta$, $\alpha$, $\zeta_c$ and $\delta$ are the obtained coefficients of the Cherenkov light LDF function (see Table 1).

**Table 1**. The coefficients of the Lorentzian function (Eq. 3) that used to parameterize the Cherenkov light LDF using CORSIKA simulation for different primary particles, different energies and different zenith angles.

| Primary particle | Zenith angle | Energy (PeV) | Coefficients | | | | $\chi^2$ | $R^2$ |
| --- | --- | --- | --- | --- | --- | --- | --- | --- |
| | | | $\eta$ | $\zeta_c$ | $\delta$ | $\alpha$ | | |
| $\gamma$ | 0° | 1 | -3.60×10⁴ | -86.51 | 67.968 | 9.78×10⁸ | 9.33×10⁸ | 0.9864 |
| $e^-$ | | | -2.12×10⁴ | -95.036 | 65.14042 | 8.86×10⁸ | 8.49×10⁸ | 0.97829 |
| P | 20° | 5 | 2.47×10⁴ | -52.705 | 25.35848 | 39.2×10⁸ | 1.83×10⁸ | 0.98112 |
| Fe | 30° | 20 | -3.21×10⁴ | -163.40 | 149.92 | 67.7×10⁸ | 3.89×10⁸ | 0.9716 |

**Results and Discussion**

The simulation of the Cherenkov radiation LDF was performed using the world-renowned CORSIKA (**CO**smic **R**ay **SI**mulation for **KA**scade) program, a detailed program of the Monte Carlo Method for the Study of the evolution and properties of charged particles in the atmosphere, which was developed to perform simulation for KASCADE experiment in Karlsruhe, Germany [5, 6]. In the current work, the LDF of Cherenkov radiation of the charged particles using the CORSIKA program version 5.61 was simulated according to the conditions and configurations of the Yakutsk EAS array. Comparative work was done to CORSIKA simulation with the Yakutsk experimental results and showed a good approximation at high energies and distances. Lorentzian function model have been used for Cherenkov light LDF of CORSIKA simulation curves as a function of the distance from the showers core in EAS. In Figure (1) it was shown the simulated Cherenkov light LDF using CORSIKA code (solid line) and that parameterized using Eq. 3 (symbol line) for gamma quanta and primary electron for vertical EAS showers at the energy 1 PeV. While in Figure (2) one can see the characteristics of primary proton and iron nuclei for inclined showers ($20^o$ and $30^o$) for different primary energies 5 PeV and 20 PeV respectively.

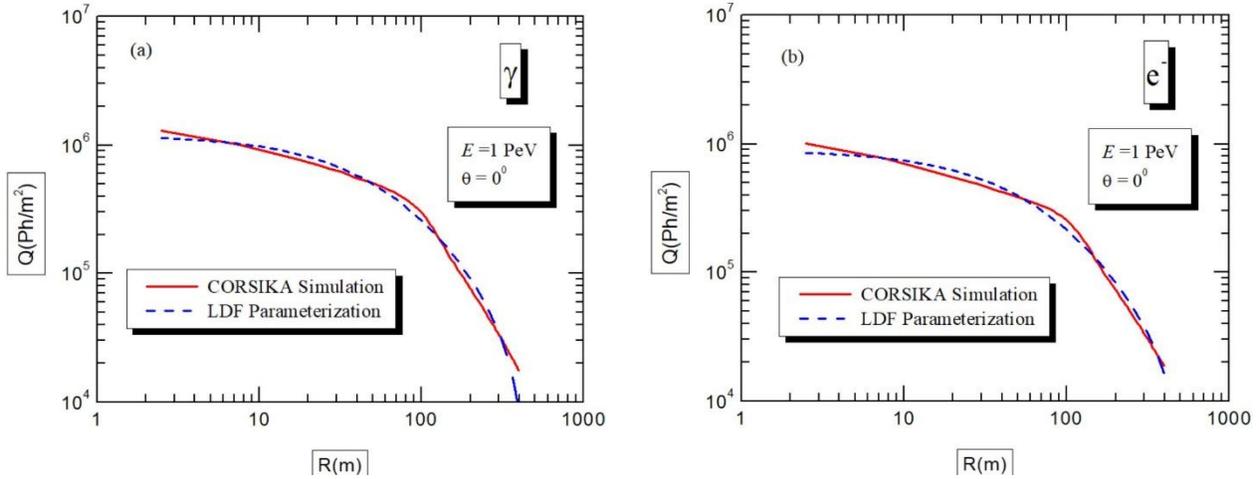

**Figure 1.** Lateral distribution of Cherenkov light which simulated with CORSIKA code (solid lines) and one parameterized (Eq. (3)) (symbol lines) at 1 PeV for vertical showers initiated (a) by primary gamma quanta; (b) by primary electron.

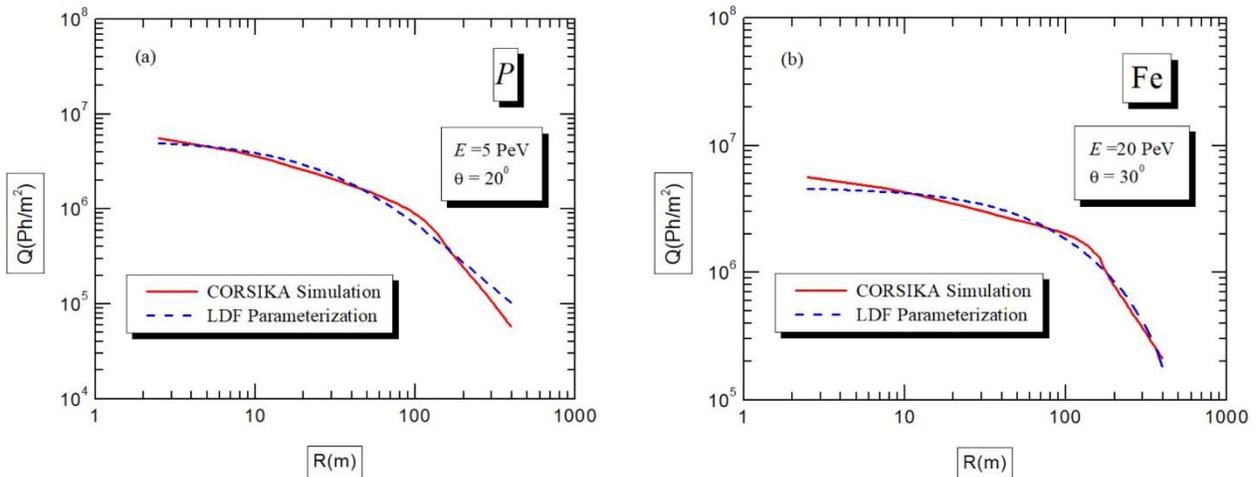

**Figure 2.** Lateral distribution of Cherenkov light which simulated with CORSIKA code (solid lines) and that parameterized (Eq. (3)) (symbol lines) at 1 PeV for inclined showers initiated (a) by primary proton at 5 PeV; (b) by iron nuclei at 20 PeV.

In Figure 3 the comparison between the Cherenkov light LDF which was calculated with Eq. (3) with that measured with Yakutsk EAS array is presented for different primary particles such as proton, iron nuclei, primary electron and gamma quanta for vertical showers.

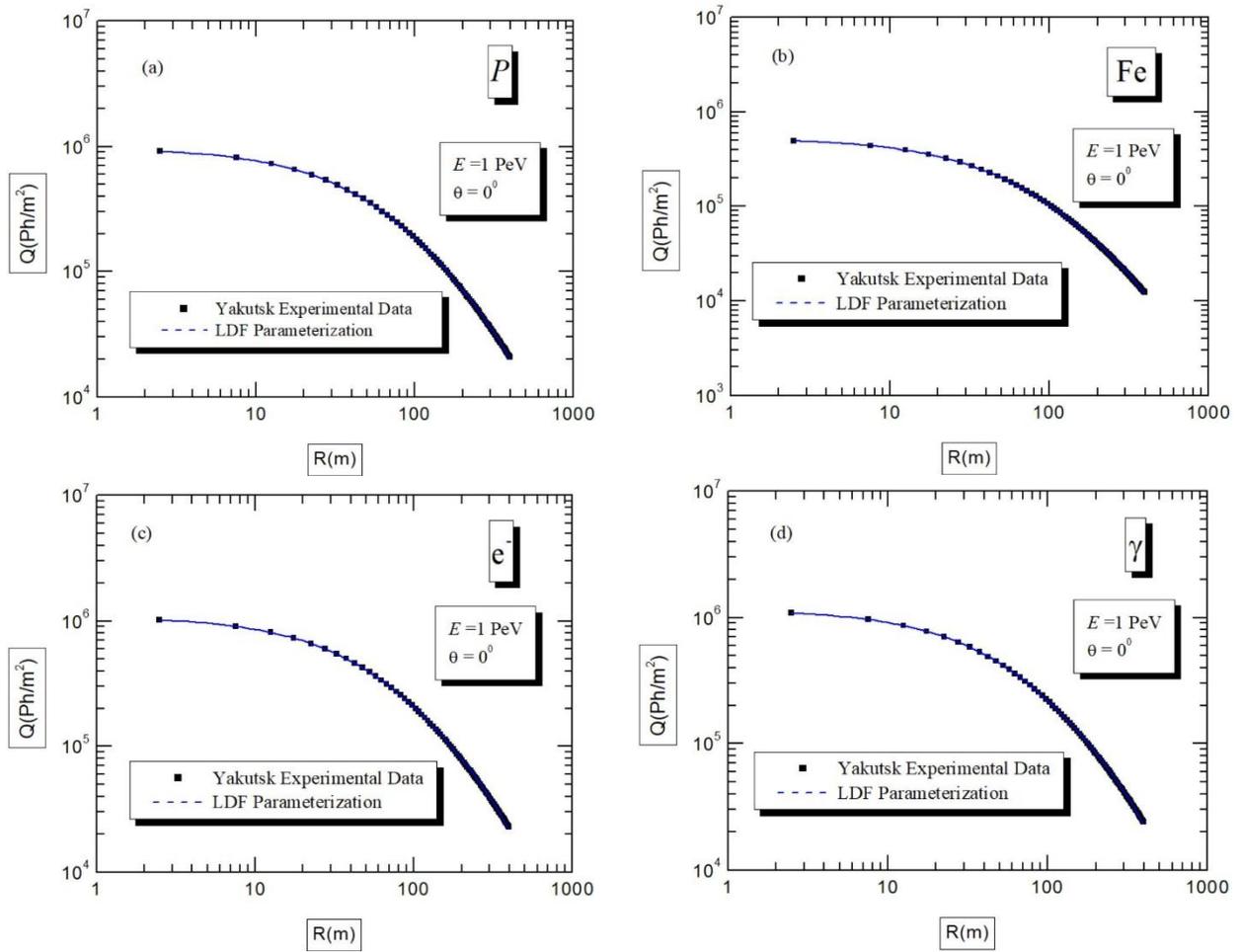

**Figure 3.** Comparison of the parameterized Cherenkov light LDF (Eq. (3)) with the data obtained by the Yakutsk EAS array (symbols) for vertical showers of (a) primary proton; (b) iron nuclei; (c) primary electron; (d) gamma quanta.

**Conclusions**

In this work the calculations of the Cherenkov light LDF in EAS initiated by different primary particles such as protons, iron nuclei, electrons and gamma quanta, were performed in the high energy range (1-20) PeV. The CORSIKA simulation of the Cherenkov light LDF in EAS is performed for configuration of the Yakutsk EAS array. The parameters of Cherenkov light LDF were obtained using the results of this simulation as a function of the distance from the shower axis. The comparison of the parameterized Cherenkov light LDF with that measured with the Yakutsk EAS array demonstrates the ability for identifying the primary particles and to determine their energies around the knee of cosmic ray spectrum. The main advantage of the given approach consists of the possibility to make a library of LDF samples which could be utilized for analysis of real events which detected with the EAS array and reconstruction of the primary cosmic rays energy spectrum and mass composition.

**Financial Disclosure:** There is no financial disclosure.
**Conflict of Interest:** None to declare.
**Ethical Clearance:** All experimental protocols were approved under the Department of Physics/ College of Science/ Mustansiriyah University, Baghdad, Iraq and all experiments were carried out in accordance with approved guidelines.